\documentclass[aps,prb,reprint,superscriptaddress,amsmath,amssymb]{revtex4-2}
\usepackage{xcolor}
\usepackage{graphicx}
\DeclareGraphicsExtensions{.png,.pdf,.tif,.jpeg}

\begin{document}


\title{Evidence of Ferromagnetic Clusters and Griffiths Singularity in Magnetic Weyl Semimetal Co$ _{3} $Sn$ _{2} $S$ _{2} $}


\author{V. Nagpal}
\email{vipin91nagpal@gmail.com}
\affiliation{School of Physical Sciences, Jawaharlal Nehru University, New Delhi-110067, India}
\author{S. Chaudhary}
\affiliation{School of Physical Sciences, Jawaharlal Nehru University, New Delhi-110067, India}
\author{P. Kumar}
\affiliation{School of Physical Sciences, Jawaharlal Nehru University, New Delhi-110067, India}
\author{Sudesh}
\affiliation{Banasthali Vidyapith University, Department of Physics, Jaipur, India 304022}
\author{S. Patnaik}
\email{spatnaik@jnu.ac.in}
\affiliation{School of Physical Sciences, Jawaharlal Nehru University, New Delhi-110067, India}


\date{\today}

\begin{abstract}
Cobalt based sulphides of compositional formula Co$ _{3} $A$ _{2} $S$ _{2} $ (A = Sn and In) are endowed with frustrated kagome lattice structure and a plethora of novel phenomena determined from the topological band structure. Here-in, we report on the detailed exploration of anisotropic magnetic properties of single crystals of ferromagnetic compound Co$ _{3} $Sn$ _{2} $S$ _{2} $. 
A low temperature clustered-glassy magnetic behaviour is revealed in field-cooled and zero field-cooled magnetization and memory effect measurement protocols. The sharp downturn and non-linearity observed in the inverse susceptibility above the critical temperature $T _{\text{C}}  $ in the paramagnetic region corroborates to the presence of short-range ferromagnetic clusters above $  T _{\text{C}}  $ in Co$ _{3} $Sn$ _{2} $S$ _{2} $. The deviation from linear Curie-Weiss behaviour in the paramagnetic state signifies the strong Griffiths singularity in the material. The slow spin dynamics behaviour and zero spontaneous magnetization above $  T _{\text{C}}  $ give an evidence of Griffiths phase owing to the ferromagnetic clusters. 
 The magnetic hysteresis loops represent the magnetization reversal which, in turn, also indicate the short range magnetic correlations, and 
reflect the coexistence of hard and soft magnetic phases in Co$ _{3} $Sn$ _{2} $S$ _{2} $. The Arrott plots derived from magnetization reveal convex type curvature at low fields and linear positive behaviour in the high field region, confirming the second order magnetic phase transition in Co$ _{3} $Sn$ _{2} $S$ _{2} $. The Takahashi spin fluctuation theory analysis provides a sufficient evidence for itinerant ferromagnetism in Co$ _{3} $Sn$ _{2} $S$ _{2} $. A large magneto-crystalline anisotropy concomitant with a high anisotropy field suggests the dominance of strong spin orbit coupling phenomenon. 
Our experimental results emphasize an intuitive understanding of the complex nature of magnetism present in Co-based shandite systems.
\begin{description}
\item[PACS numbers]
\item[Keywords]shandite; ferromagnetic clusters; Griffiths singularity; susceptibility; hysteresis; anisotropy; itinerant; magnetocrystalline; spin-orbit coupling
\end{description}
\end{abstract}


\maketitle


\section{\label{sec:level1}Introduction}
Cobalt based sulphides with formula Co$ _{3} $A$ _{2} $S$ _{2} $ (A = Sn and In) are considered representative examples of strongly correlated $ d $-electron systems that have triggered immense interest for their remarkable physical properties \citep{1,2,3,4,5,6}. In particular, the coexistence of half-metallic state and ferromagnetism in shandite-type Co$ _{3} $Sn$ _{2} $S$ _{2} $ makes it promising for spintronics and technological applications \cite{7,8}. Reported band structure calculations on Co$ _{3} $Sn$ _{2} $S$ _{2} $ reveal predominantly Co 3$ d $-character in addition to Sn 5$ s $ and S 3$ p $ states. This results in sharp, narrow bands in the vicinity of Fermi level $E_{\text{F}}$ leading to the low dimensional character of layered kagome structure \citep{3,4,5,6,7,8,9}. Co$ _{3} $Sn$ _{2} $S$ _{2} $ attains a ferromagnetic state below $T _{\text{C}} =177 $ K.  It is characterized as a type $ I_{\text{A}} $ half-metallic ferromagnet from photoemission and transport measurements \citep{8,9,10}. Furthermore, Angle Resolved Photoemission Spectroscopy (ARPES) and Scanning Tunnelling Microscopy (STM) experiments confirm the Weyl semimetal phase in Co$ _{3} $Sn$ _{2} $S$ _{2} $. Consequent phenomena such as anomalous Hall effect, anomalous Nernst effect and magneto-caloric effect have also been reported \citep{11,12,13,14}. 
\par Several recent works based on Co-based shandite minerals have offered analysis of structural, magnetic and electronic phases, that exhibit spin glass state and metal-semiconductor phase transition owing to the geometric frustration in the kagome arrangement of magnetic Co atoms \cite{15,16}. These inhomogeneous phases originate from the slow spin dynamics and shifting of the Fermi level to lower occupied energy states below the band gap \cite{16}. Moreover, these phase inhomogeneities in the magnetic materials are associated with formation of ferromagnetic clusters (FMC) embedded in the paramagnetic region \citep{17,18,19}. In other words, the short range FMC occur at a higher temperature $T _{\text{GP}}$ far above the long-range magnetic ordering temperature $T _{\text{C}} $, which is primarily ascribed to Griffiths singularity. The region where FMC exists, i.e., between the temperatures $T _{\text{C}} $ and $T_{\text{GP}}$, characterizes the Griffiths phase (GP). In GP region, disordered finite size clusters with uncorrelated spins exist and a non-analytic behaviour of magnetization is reported \cite{17,18}. This phenomenon was first identified in randomly diluted Ising ferromagnets \cite{20}. Such FMC have been observed in complex oxides where identified reasons include quenched chemical or structural disorder, or spin fluctuations \cite{21,22}.
\par The present work primarily focuses on comprehensive investigation of the magnetic properties of shandite Co$ _{3} $Sn$ _{2} $S$ _{2} $. The DC magnetization measurement shows the evidence of low temperature glassy state.  The divergence from Curie-Weiss behaviour above the transition temperature is indicative of Griffiths-like phase. The magnetic field dependent magnetization exhibits a hysteresis with low coercivity. The high magneto-crystalline anisotropy is determined using law of approach to saturation method.  We also analyse our magnetic data on the basis of Takahashi spin fluctuation theory which explains the physical properties of itinerant ferromagnetic materials in electron correlated systems \cite{23}. 
\section{\label{sec:level2}Experiment}
Single crystals of Co$ _{3} $Sn$ _{2} $S$ _{2} $ were prepared from stoichiometric amounts of Co (Sigma Aldrich, 99.999\%), Sn (Sigma-Aldrich, 99.999\%) and S (Sigma-Aldrich, 99.999\%). The samples were synthesized using modified Bridgmann technique, as described in our previous article \cite{24}. The phase purity and crystal structure of the as-synthesized crystals was confirmed from X-ray diffraction (XRD) using Rigaku Miniflex 600 instrument. The magnetization measurements on the samples were performed using Vibrating Sample Magnetometer (VSM) installed in \textit{Cryogenic} Physical Properties Measurement System (PPMS). 	

\section{\label{sec:level3}Results and Discussions}
Figure 1(a) shows the XRD pattern of the cleaved surface of Co$ _{3} $Sn$ _{2} $S$ _{2} $ single crystal. The $00l$ peaks in the diffraction pattern suggests the single crystalline nature of the sample with the cleaved surface of crystal being the basal $ab$ plane. The Rietveld refinement analysis of powder XRD pattern of  Co$ _{3} $Sn$ _{2} $S$ _{2} $ confirmed the phase purity and the refined lattice parameters were obtained to be $a = 5.37$ \AA{} and $c = 13.17$ \AA{} \cite{24}. The shandite structure of Co$ _{3} $Sn$ _{2} $S$ _{2} $ constitutes quasi two-dimensional (2D) sheets of magnetic Co atoms in a hexagonal arrangement forming a kagome-type lattice. These sheets are capped above or below by S atoms, and stacked in an ABC-ABC chain sequence (figures 1(b) and 1(c)). The Sn atoms coordinated between the kagome sheets are coupled to trigonal anti-prismatic interlayer sites generated by triangular layers of adjacent Co atoms. The Co and S atoms occupy the unique lattice sites 9e (1/2,0,0) and 6c (0,0,z) respectively whereas Sn atoms are partially located at two different sites, 3b (0,0,1/2) and 3a (0,0,0).
\begin{figure}
\includegraphics[scale=0.35]{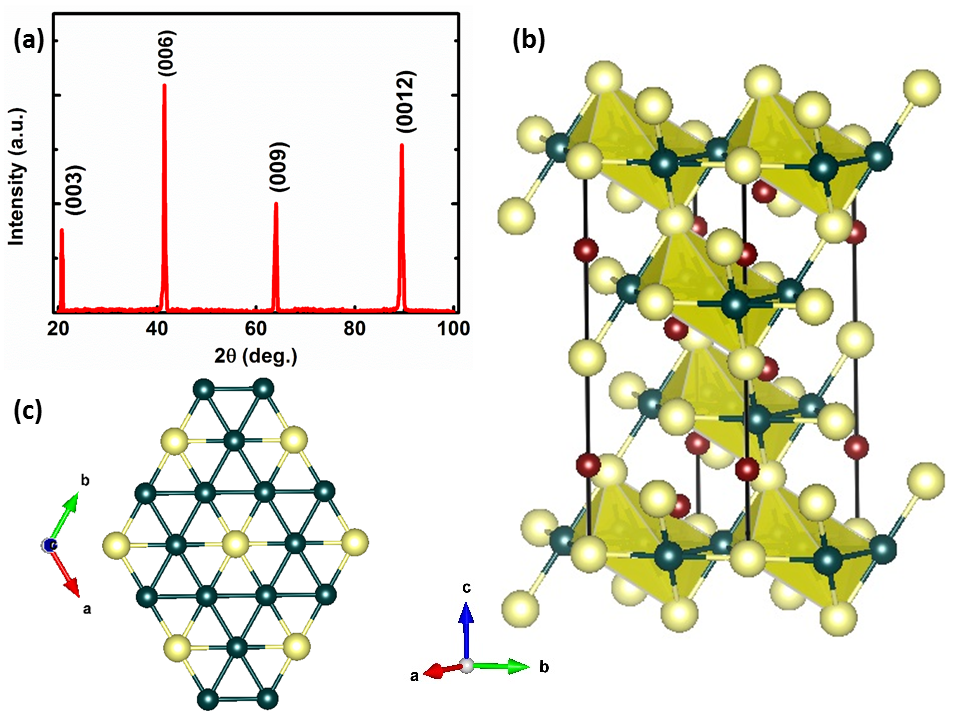}    
 \caption{
 (a) XRD pattern of the cleaved surface i.e., ($ 00l $) plane of Co$ _{3} $Sn$ _{2} $S$ _{2} $ single crystal. (b) Shandite crystal structure of Co$ _{3} $Sn$ _{2} $S$ _{2} $ compounds. Atoms Co, Sn and S are represented by green, light orange and magenta colours. Co atoms centred in trigonal anti-prismatic layers forming a kagome network with Sn atoms (yellow polyhedra). (c) Hexagonal arrangement of Co atoms.} 
\end{figure}
\begin{figure*}
\begin{minipage}{1.0\textwidth}
\includegraphics[width=\textwidth]{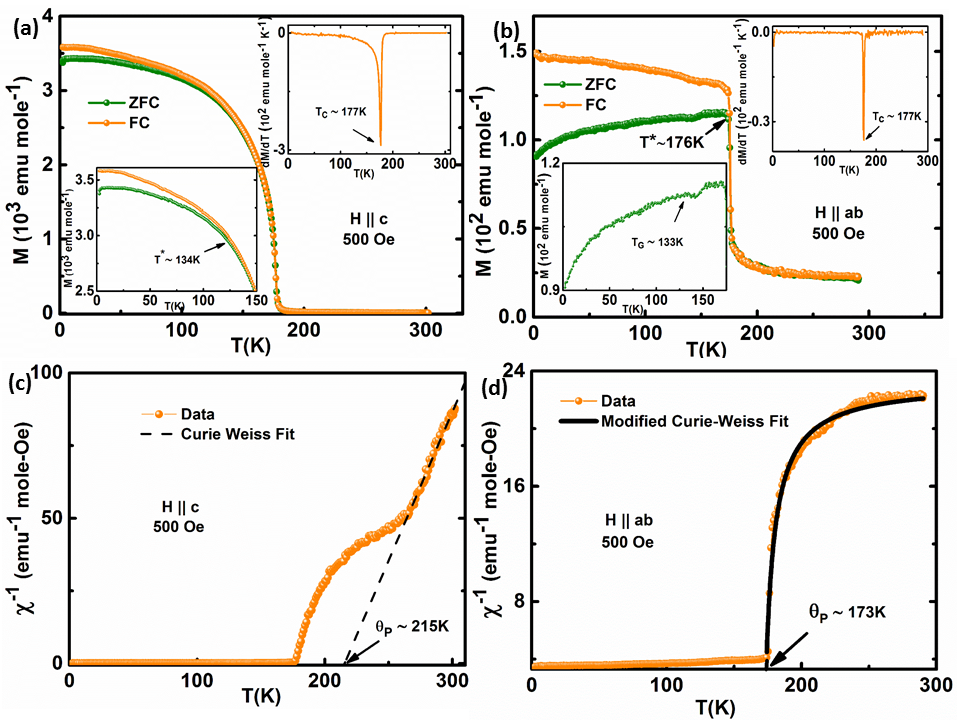}  
\caption{Temperature dependent zero-field cooled (ZFC) and field-cooled (FC) magnetization $ M(T) $ measurements taken at an external magnetic field $ H $ of 500 Oe in (a) $H\parallel c$  and (b) $H\parallel ab$ directions. The upper insets in (a) and (b) show $ dM/dT $  vs. $ T $. The minimum in $ dM/dT $ plot represents the transition temperature $ T _{\text{C}} $. The lower inset in (a) show the FC-ZFC bifurcation in the low temperature magnetization data marked at $ T^{\ast} = 134 $ K for $H\parallel c$ direction. The lower inset in (b) depicts the ZFC curve where the peak marked at $ T_{\text{G}} \approx 133 $ K represents the spin glass transition in $H\parallel ab$ direction. Inverse magnetic susceptibility $ \chi^{-1} $ as a function of temperature is shown in (c) $H\parallel c$ and (d) $H\parallel ab$ directions. The inverse susceptibility in (c) is fitted with linear Curie-Weiss equation: $\chi=C/(T-\theta_{\text{P}})$ above 250 K and (d) is fitted with modified CW equation: $\chi=\chi_{0}+C/(T-\theta_{\text{P}})$. The fitting parameters $  C $ and $ \theta_{\text{P}} $ give the effective magnetic moment and Curie temperature, respectively.}
\end{minipage}
\end{figure*}
\par The temperature dependent DC magnetization $M(T)$ in zero field cooled (ZFC) and field cooled (FC) modes, under an applied magnetic field $H=500$ Oe, performed in both out-of-plane ($H\parallel c$-axis) and in-plane ($H\parallel ab$ plane) directions are shown in figures 2(a) and 2(b), respectively. The respective differential of magnetization $ dM/dT $ vs. $ T $ (inset in figures 2(a) and 2(b)) plots exhibit minima around 177 K, confirming the ferromagnetic (FM) to paramagnetic (PM) phase transition. Interestingly, there is a significant difference observed in $M(T)$ data below $ T _{\text{C}}  $ in two crystallographic directions implying the anisotropic nature of Co$ _{3} $Sn$ _{2} $S$ _{2} $ \cite{8}. Similar results have also been observed previously in Co$ _{3} $Sn$ _{2} $S$ _{2} $ where the presence of strong magnetic anisotropy was ascribed for this appreciable different behaviours at field parallel to c-axis and ab-plane \cite{10,14}.
Based on earlier discussions on two dimensional layered kagoome ferromagnets \cite{25,26,27}, the strong magnetic anisotropy in Co$ _{3} $Sn$ _{2} $S$ _{2} $ is suggested to predominantly originate from the combined effects of various quantum phenomena such as strong spin-orbit coupling \cite{28,29}, crystal electric field \cite{29}, exchange couplings \cite{30}, and stripe-like features of ferromagnetic domains\cite{31}. Besides, the difference in magnetic behaviour along the two field directions can also possibly be associated with the presence of large magnetocrystalline anisotropy energy caused by the intrinsic strong spin-orbit coupling \cite{28}. In general, the anisotropic behaviour in a magnetic material is believed to originate from the intrinsic atomic magnetic moments associated with magnetocrystalline anisotropy \cite{31}. As a consequence of magnetocrystalline anisotropy, the anisotropy in magnetization is observed in our measurement data \cite{32}.
\par Furthermore, the FC-ZFC curves in both the directions show a bifurcation below the critical temperature $ T _{\text{C}}  $. In general, this can be ascribed to magnetic anisotropy, formation of FM clusters, spin frustrations or pinning of the ferromagnetic domain walls 
\citep{33,34,35,36,37}. It is clearly seen that the bifurcation onsets at higher temperature $ T^{\ast} \sim 176 $ K in $H\parallel ab$ direction (figure 2(b) main panel) as compared to $ T^{\ast} \sim 134 $K in $H\parallel c$  direction (figure 2(a) lower inset). This, in turn, implies that the compound is highly anisotropic where the spins get locked in random directions, resulting in decrease in magnetization at low temperatures \cite{38}. In $H\parallel ab$ direction, a peak is also detected at $ T_{\text{G}} \sim 133 $ K in ZFC curve marked by an arrow in the lower inset of figure 2(b). This is an indication of frozen spin states depicting the low temperature glassy transition \cite{16}. These features such as FC-ZFC bifurcation and peak in ZFC curve are typical characteristics of spin glass state \citep{39,40,41}. Such scenario has also been reported in systems involving cluster glasses \cite{40}, super-paramagnets \cite{42}, and superconducting oxides \cite{43}.
\par Next, we discuss the magnetic behaviour of Co$ _{3} $Sn$ _{2} $S$ _{2} $ using the temperature dependent susceptibility $\chi(=M/H)$ as derived from the magnetization measurements. It is quite familiar that the susceptibility as a function of temperature in the PM region follows Curie-Weiss (CW) law: $\chi=C/(T-\theta_{\text{P}})$, where $\theta_{\text{P}}$ is the Weiss temperature and $C$ is the Curie constant. The effective magnetic moment $ \mu_{\text{eff}} $ is calculated from the Curie constant using the expression: 
\begin{equation}
\mu_{\text{eff}}=\sqrt{\frac{3k_{\text{B}}C}{N_{\text{A}}}}
\end{equation}$  $
where $ k_{\text{B}} $ is the Boltzmann constant and $ N_{\text{A}} $ is the Avogadro number.
\par Figures 2(c) and 2(d) illustrate the inverse of magnetic susceptibility $\chi^{-1}$ as a function of temperature in the applied magnetic field 500 Oe in $H\parallel c$ and $H\parallel ab$ directions, respectively. The plots of $\chi^{-1}$  vs. $ T $ above the magnetic transition $T _{\text{C}}$ exhibit a strong curvilinear behaviour in both directions. In addition, the distinct behaviour in inverse magnetic susceptibility data is observed along the two directions that is likely due to the strong magnetocrystalline anisotropy where $\chi^{-1}$ data along c-axis is four times larger than that in ab-plane \cite{31,32}. An unusual divergence in the PM region from CW linear behaviour is observed above $T _{\text{C}}  $ 
in the $H\parallel c$ direction. In contrast, a non-linear concave curvature is observed in $H\parallel ab$ direction above $T _{\text{C}}  $ that continues upto room temperature. Consequently, a sharp downturn is detected in both the directions as the temperature is decreased to $T _{\text{C}}  $ from the paramagnetic state. This non-analytic behaviour of magnetization is indicative of the emergence of short-range FM clusters before the long range FM ordered state sets below $T _{\text{C}}  $ \cite{17,18}. It is suggested that the deviation from linear CW behaviour in such plots is a consequence of exchange splitting between Co-3$ d $ orbital states due to the spin-orbit coupling \cite{44}. Other factors including phase inhomogeneity, quenched  disorder and spin fluctuations have also been discussed \cite{21,22}. Moreover, the downturn in inverse susceptibilities is considered to be a characteristic signature of Griffiths phase, which is different from a smeared phase transition that gives rise to an upward deviation in $\chi^{-1}$ in the PM state \cite{20}. Similar features in inverse susceptibility have been reported in semiconducting ferromagnet La$ _{1-x} $Sr$ _{1+x} $CoO$ _{4} $ $ (0\leqslant x \leqslant1) $, half doped manganite Pr$ _{0.5} $Sr$ _{0.5} $MnO$ _{3} $, and La-based manganites \cite{21,22,45}.
The plot of $\chi^{-1}$  vs. $ T $ for $H\parallel c$ direction above 265 K is found to be perfectly in agreement with CW fitting, (figure 2(c)). The slope and intercept parameters of linear CW fitting give the calculated $ C $ and $ \theta_{\text{P}} $, respectively. In contrast, the $\chi^{-1}$  vs. $ T $ plot in $H\parallel ab$ above $T _{\text{C}}  $ does not follow the conventional linear CW law and hence, the concave curvature, (figure 2(d)), is fitted with a modified Curie-Weiss equation: $\chi=\chi_{0}+C/(T-\theta_{\text{P}})$ where $ \chi_{0} $ is the temperature independent factor arising from Pauli paramagnetic and diamagnetic contributions \citep{44,46}. Using the fitted curves, the obtained values of $ \theta_{\text{P}} $ in both the cases are positive, and found to be 215 K and 173 K respectively, validating the dominance of ferromagnetic interactions among the spins. Similar results of $ \theta_{\text{P}} $ values determined away from $T _{\text{C}}  $ were also reported in perovskite and mixed valence manganites \cite{47,48}. In general, the difference in the values of $ \theta_{P} $ and $T _{\text{C}}  $ i.e, $ \theta_{\text{P}} > T _{\text{C}}  $ or $ \theta_{\text{P}} < T _{\text{C}}  $ purely depends on the inhomogeneous phase present in the substance, which in turn is linked to the presence of short-range magnetic correlations just above $T _{\text{C}}  $ \cite{49}. In $H\parallel c$ direction, $\mu_{\text{eff}}$ is determined to be 2.77$ \mu_{\text{B}} $ from Eq. (1). The obtained $\mu_{\text{eff}}$ is intermediate between the theoretically calculated (spin-only) low-spin state (1.73$ \mu_{\text{B}} $) and high-spin state (3.87$ \mu_{\text{B}} $) values of Co$ ^{2+} $ ions \cite{50}. On the other hand, the experimental value of $\mu_{\text{eff}}$ for $H\parallel ab$ direction is 6.63$ \mu_{\text{B}} $ using the modified Curie-Weiss fit.
The difference in the results of experimental and theoretical values of $\mu_{\text{eff}}$ in both the directions supports the development of short range FM clusters near $T _{\text{C}}  $ \cite{51}. Further, the high value of $\mu_{\text{eff}}$ means a contribution of both orbital and spin components to the magnetic moment due to the large splitting of orbital levels relative to thermal energy ($ k_{\text{B}}T $) \cite{52}. This high $\mu_{\text{eff}}$ value further confirms the FM correlations present in the PM regime. 
\begin{figure}
\begin{minipage}{0.5\textwidth}
\includegraphics[width=\textwidth]{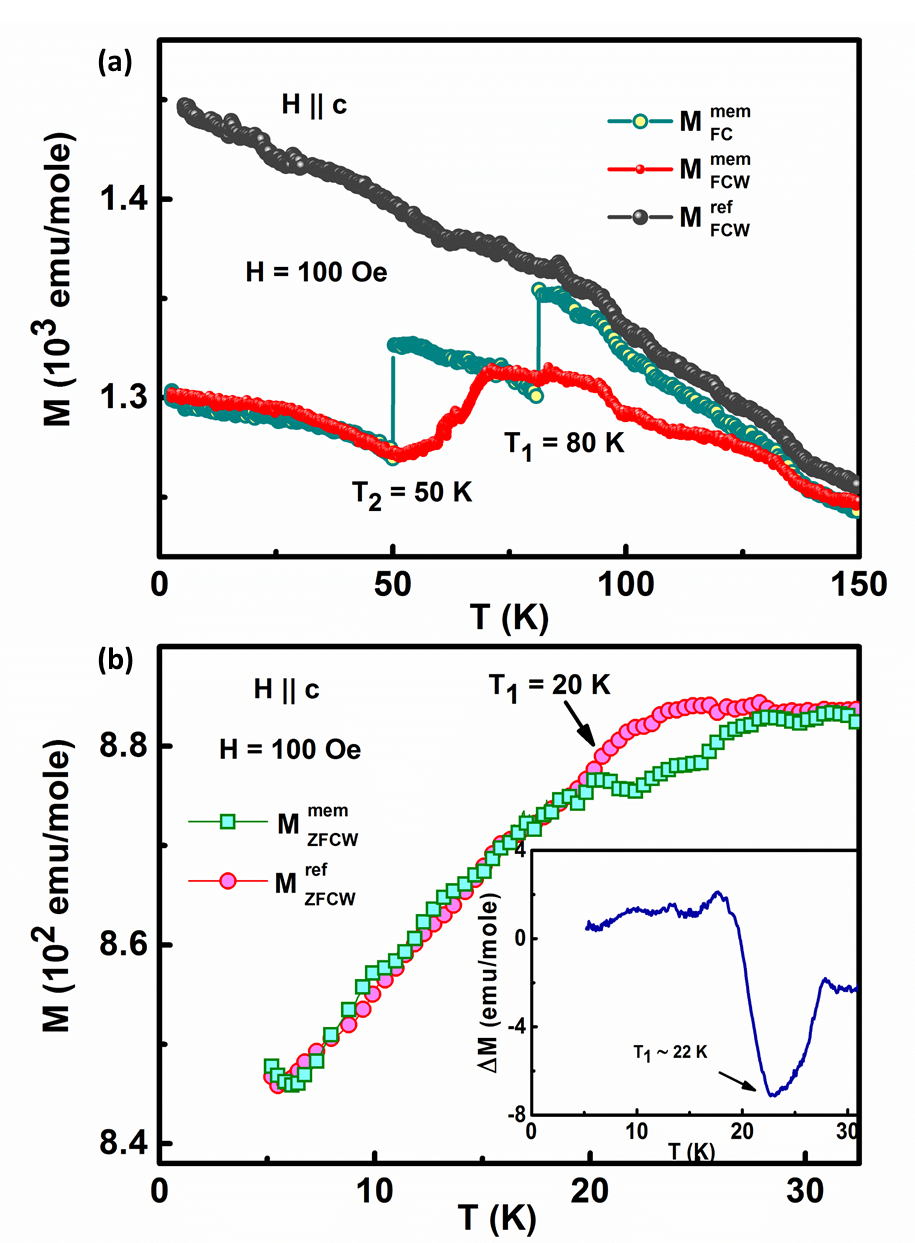} 
\end{minipage}
\caption{The variation of memory effect with temperature in (a) FC and (b) ZFC modes under the applied magnetic field $ H = 100 $Oe. The processes are interrupted for the period of two hours each at $ T_{1}=80 $ K and $ T_{2}=50 $ K in FC mode and $ T_{1}=20 $ K in ZFC mode. Inset in (b) shows the difference $ \Delta M = M^\text{ref}_\text{ZFCW} - M^\text{mem}_\text{ZFCW} $ as the function of temperature in ZFC mode.}
\end{figure}

\par In order to ascertain the nature of low temperature glassy state in Co$ _{3} $Sn$ _{2} $S$ _{2} $, the memory effect measurements were carried out in both FC and ZFC modes in $H\parallel c$ direction as shown in figures 3(a) and 3(b), respectively \cite{53,54}. Firstly, the sample was treated under FC mode while cooling from room temperature to 5 K in the presence of 100 Oe applied magnetic field. The FC process was interrupted for a duration of 2 hours each at $ T_{1}=80 $ K and $ T_{2}=50 $ K below the glassy transition $ T_\text{G}$ where the magnetic field was turned off allowing the system to acquire a relaxed state. At each temperatures after the waiting period was completed, the same magnetic field was turned on and the cooling process was resumed. The magnetization data recorded using this process is referred to as $ M^\text{stop}_\text{FC} $ which shows step-like behaviours at the interrupted temperatures 80 K and 50 K. After cooling down to 5 K, the sample was heated up to 300 K under the same magnetic field without any interruption and the obtained magnetization curve is denoted as $ M^\text{mem}_\text{FCW} $. It is interesting to note that $ M^\text{mem}_\text{FCW} $ curve also exhibits the characteristic dips at each halt temperatures performed in $ M^\text{stop}_\text{FC} $, thus following the previous $ M(T) $ data. Hence, this feature clear indicates the magnetic memory effect in our sample. Again, the FC magnetization $ M^\text{ref}_\text{FCW} $ is taken for reference in the same field without any interruption. Similarly, the memory effect measurement was also performed in ZFC condition along the same direction in which the sample was cooled down from room temperature to 5 K in the absence of magnetic field. The ZFC process was interrupted at $ T_{1} = 20 $ K for two hours duration where the magnetic field was turned off. As soon as the temperature reaches 2 K, a 100 Oe magnetic field was applied and the magnetization data was measured while warming the sample which is referred to as $ M^\text{mem}_\text{ZFCW} $. Again, the sample is cooled down to 2 K in zero magnetic field without any interruption and the magnetization data was recorded while warming the sample in the applied magnetic field of 100 Oe which is denoted as $ M^\text{ref}_\text{ZFCW} $. It is clearly noted the $ M^\text{mem}_\text{ZFCW} $ and $ M^\text{ref}_\text{ZFCW} $ curves coincides each other except around 20 K. Clearly, a difference in  i.e., a memory dip at 22 K close to interrupted temperature is observed as a result of the difference between $ M^\text{mem}_\text{ZFCW} $ and $ M^\text{ref}_\text{ZFCW} $ curves i.e., $ \Delta M = M^\text{ref}_\text{ZFCW} - M^\text{mem}_\text{ZFCW} $  as shown in the inset of figure 3(b). Since the ZFC memory effect is observed only in glassy magnetic systems and not in superparamagnetic systems even if both exhibit FC memory effect, one can explicitly rule out the condition of superparamagnetism for the observation of FC and ZFC memory effect \cite{55}. Hence, we strongly conclude from memory effect results that the origin of glass-like features at low temperatures is due to the presence of ferromagnetic clusters of spins present in this system \cite{53}.
\begin{figure}
\begin{minipage}{0.45\textwidth}
\includegraphics[width=\textwidth]{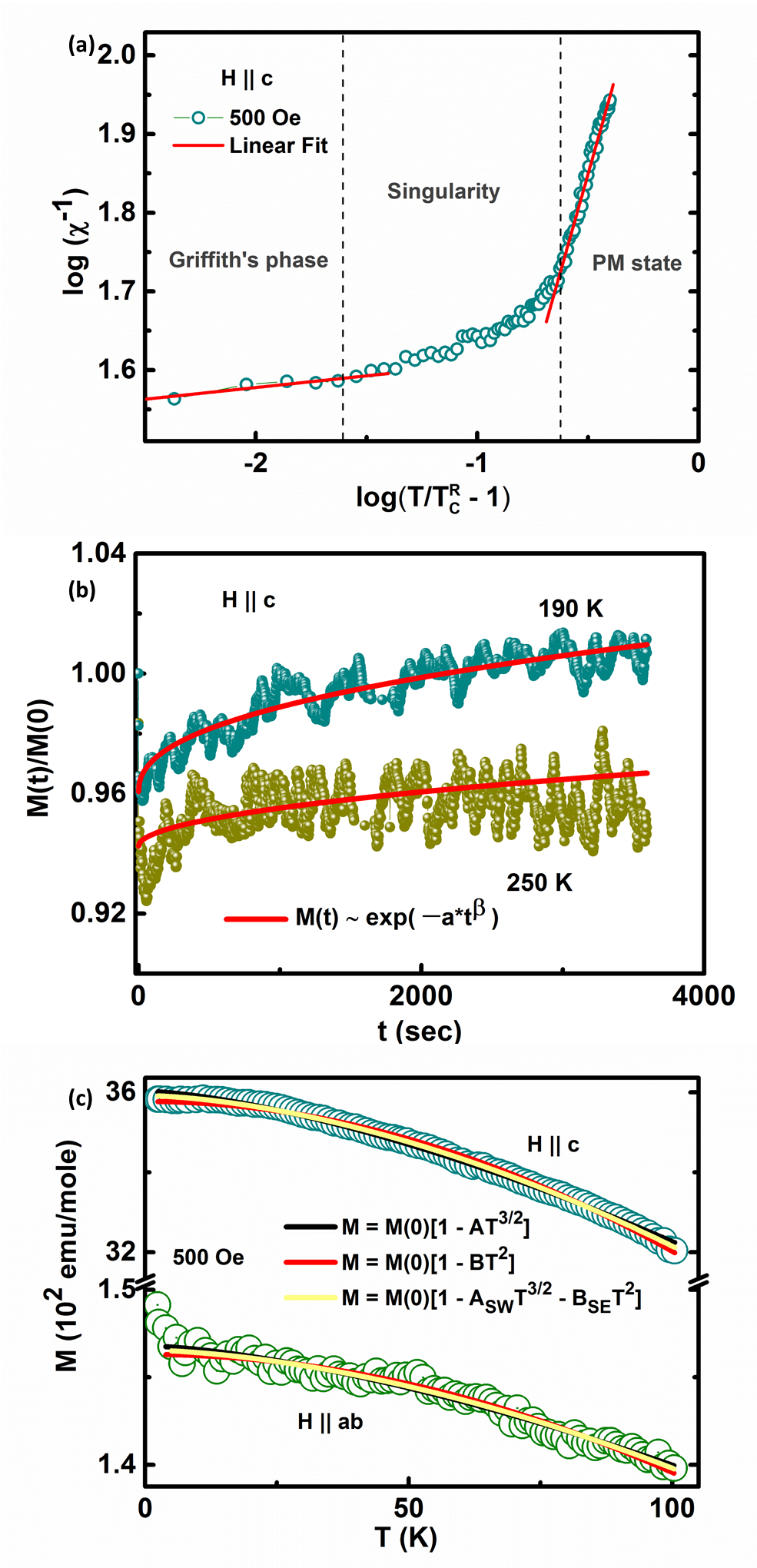}  
 \caption{(a) Plot of $ \log (\chi^{-1}) $ vs. $ \log(T/T_\text{C}^\text{R} - 1) $ for $H\parallel c$ direction according to Eq. (1). The red solid line is the linear fitting to the derived data. (b) The normalized magnetization $ M(t)/M(0) $ with the evolution of time at 190 K and 250 K fitted with an equation: $ M(t)\sim exp(-at^{\beta}) $ (red solid curve). (c) Temperature dependence of magnetization measured in the low temperature range (2-100 K) for both $H\parallel c$ and $H\parallel ab$ directions. The black solid line is the fit according to equation: $M(T)=M(0)[1-A_{\text{SW}} T^{3/2}] $. The red solid line is the fit according to equation: $ M(T)=M(0)[1-B_{\text{SE}} T^{2}] $. The yellow line is the fit for the equation: $ M(T)=M(0)[1-A_{\text{SW}} T^{3/2}-B_{\text{SE}} T^{2}] $.} 
\end{minipage}
\end{figure}
\par As stated earlier, the curved downturn in the reciprocal magnetic susceptibility above the FM phase transition is a typical feature of Griffiths phase characterised by the appearance of finite-size FM clusters with the spin correlations in the PM matrix \cite{17,18}. The GP phase is a clustered state that possess local spin interactions in the PM region. In this GP region, the spins are ferromagnetically correlated within the clusters and the system is expected to exist in neither long range FM ordered state nor pure PM state. As a consequence, no spontaneous magnetization would appear in this GP region and hence, magnetization shows non-analytic behaviour. This observation arises from the Griffiths singularity which causes the susceptibility deviation from CW behaviour \cite{20}. Based on the theoretical models, the divergence in susceptibility in Griffiths phase leads to a power-law behaviour expressed as \cite{22,56}:
\begin{equation}
\chi^{-1}\propto(T-T_\text{C}^\text{R})^{1-\lambda}
\end{equation}
Here, $ T_\text{C}^\text{R} $ refers to the critical temperature of random FM at which susceptibility tend to diverge and the exponent $ \lambda (0 < \lambda \leqslant 1) $ characterises the Griffiths singularity which signifies the deviation from CW behaviour. The power-law relation in Eq. (2) is a modified form of CW law. The value of exponent $ \lambda $ tends to be zero in the pure PM region. So, it is understood that the high value of $ \lambda $ implies strong deviation from CW behaviour.
\begin{figure*}
\includegraphics[width=\textwidth]{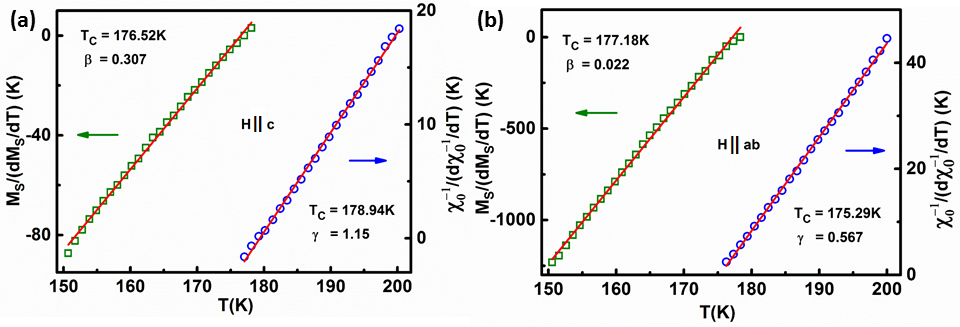}     
 \caption{Kouvel-Fisher (KF) plots of Co$ _{3} $Sn$ _{2} $S$ _{2} $ for spontaneous magnetization (Eq. (3)) on the left axis and inverse susceptibility (Eq. (4)) on the right axis in (a) $H\parallel c$  and (b) $H\parallel ab$ directions. The red solid line is a linear fit to the extracted data.} 
\end{figure*}
\par In order to calculate the strength of deviation from CW law, the magnetic susceptibility is plotted as $\chi^{-1}$ vs. $ (T/T_\text{C}^\text{R} - 1) $ at 500 Oe on log-log scale for $H\parallel c$ direction (figure 4(a)). The slopes obtained from linear fitting in the GP regime and PM region give the respective values of exponents $ \lambda_{\text{GP}} $ and $ \lambda_{\text{PM}} $. It is discussed that the proper value of $ \lambda $ must be determined from an accurate $ T_\text{C}^\text{R} $, otherwise an improper $ T_\text{C}^\text{R} $ in Eq. (2) would lead to unphysical fitting and erroneous determination of $ \lambda $. To estimate the correct $ T_\text{C}^\text{R} $, we have followed a method which initially fixes $ T_\text{C}^\text{R} $ equal to $T _{\text{C}}  $ and $ \lambda_{\text{PM}} $ is evaluated from the fitted plot. Then, the value of $ T_\text{C}^\text{R} $ is adjusted such that $ \lambda_{\text{PM}} $ comes close to zero in the PM state. It follows the fact that GP gets transformed to usual PM region above the Griffiths temperature $ T _{\text{GP}} $ and $\chi^{-1}$ obeys the CW law. Here, $ T _{\text{GP}} $ is evaluated from the starting point of downturn in $\chi^{-1}$. Using this method, the respective $ T_\text{C}^\text{R} $ and $ T _{\text{GP}} $ values are found to be 216 K and 265 K. Further, the slopes from the linear fitting in figure 4(a) gives $ \lambda_{\text{PM}} =0.003$ and $ \lambda_{\text{GP}} =0.967$. Therefore, we confirmed that a very large value of $ \lambda_{\text{GP}} $ reflects the remarkably strong Griffiths singularity in this material. 
\par The existence of Griffiths singularity is commonly characterised by the slow dynamics of spins \cite{57}. It is expected that the spin relaxation is slower due to the short range FM clusters in GP phase rather than that in PM regime. Therefore, in order to provide more evidence of the presence of short-range FM clusters, the isothermal remanent magnetization measurements were performed to study the spin relaxation behaviour in the GP phase below $ T _{\text{GP}} $. The magnetization $ M(t) $ was recorded with the evolution of time in the presence of 500 Oe magnetic field at 190 K and 250 K in the GP regime. Initially, the sample in zero magnetic field was cooled upto the desired temperature and then, the magnetization was scanned upto 3600 s after applying the magnetic field. Figure 4(b) shows the normalized magnetization $ M(t)/M(0) $ as a function of time exhibiting the slow spin dynamics where $ M(0) $ is the magnetization at $ t=0 $. It implies the spin relaxation behaviour and the magnetization is observed to follow the exponential form: $ M(t)\sim exp(-at^{\beta}); (0<\beta<1) $. Using the exponential fit, $ \beta \sim$ 0.42 is determined below $ T _{\text{GP}} $ \citep{58}. Hence, such slow relaxation behaviour at 190 K and 250 K below $ T _{\text{GP}} $ is due to the presence of the ferromagnetic clusters of uncorrelated spins \cite{58,59}.
\par Next, we discuss the low temperature thermal effects of magnetization in both the directions. It is ascertained that the low temperature magnetization in some ferromagnets commonly decreases with the increase in temperature due to thermally excited magnons and Stoner excitations \cite{60}. The magnons follow Bloch $ T^{3⁄2} $ law unlike the $ T^{2} $ dependence for Stoner excitations. Following this approximation, the measured FC magnetization in the low temperature region in $H\parallel c$ and $H\parallel ab$ directions are shown in figure 4(c). The magnetization data is fitted to (i) Bloch law: $M(T)=M(0)[1-A_{\text{SW}} T^{3/2}] $, (ii) the Stoner excitation term: $ M(T)=M(0)[1-B_{\text{SE}} T^{2}] $, and (iii) combination of Bloch and Stoner excitation terms: $ M(T)=M(0)[1-A_{\text{SW}} T^{3/2}-B_{\text{SE}} T^{2}] $. Here $ M(0)$ is the magnetization at 0 K. $ A_{\text{SW}} $ and $ B_{\text{SE}} $ are spin-wave and Stoner-excitation parameters, respectively. It is observed that the low temperature magnetization for both the directions is also separately well fitted with the Bloch function as well as the Stoner excitation term. Considering the combined fit according to (iii), the values of $ M(0) $ are calculated as $ 3.6 \times 10^{3} $ emu/mole and $ 0.14 \times 10^{3} $ emu/mole for $H\parallel c$ and $H\parallel ab$ directions, respectively. Also, the parameters $ A_{\text{SW}} $, $ B_{\text{SE}} $ are found to be $ 5.89 \times 10^{-5} \text{K}^{-3⁄2}$, $ 4.56 \times 10^{-6} \text{K}^{-2}$ and $ 2.59 \times 10^{-5} \text{K}^{-3⁄2}$, $ 2.01 \times 10^{-6} \text{K}^{-2}$ for $H\parallel c$ and $H\parallel ab$ orientations, respectively. It is noted  that the values of $ B_{\text{SE}} $ are found smaller in comparison to $ A_{\text{SW}} $ in both the directions which implies that the Stoner excitations are suppressed by the dominant spin wave excitations \cite{61}. Using the fitted spin wave parameter $ A_{\text{SW}} $, the exchange interaction $ J_{\text{ex}} $ is evaluated between two Co$ ^{2+} $ neighbouring ions from the expression: $ A_{\text{SW}}=(0.0587⁄S)(2J_{\text{ex}}S)^{3/2} $, where $ S=3⁄2 $ is the total spin of Co$ ^{2+} $ ions in the high spin state. The values of $ J_{\text{ex}} $ are found to be 25.38 $ k_{\text{B}} $K and 43.89 $ k_{\text{B}} $K respectively for $H\parallel c$ and $H\parallel ab$ directions. The difference in the values of exchange interaction in both the directions is due to the change in interatomic distance between neighbouring Co$ ^{2+} $ ions in two orientations.
We next examine the magnetic behaviour near the critical temperature region in Co$ _{3} $Sn$ _{2} $S$ _{2} $ by employing Kouvel-Fisher (KF) method where the spontaneous magnetization $ M_{S} $ and inverse susceptibility $ \chi_0^{-1} $ are evaluated using equations (3) and (4) \cite{62}. The KF method is often chosen to be more accurate compared to other methods for determining $T _{\text{C}}  $ and the critical exponents $ (\beta,\gamma) $ in magnetic systems \cite{63,64}.  The critical exponents are the characteristic magnetic parameters near the magnetic phase transition which specifies the nature of magnetic ordering and interactions involved in the system. 
\begin{equation}
\dfrac{M_{S}(T)}{dM_{S}(T)/dT}=\dfrac{(T-T _{\text{C}} )}{\beta}
\end{equation}
\begin{equation}
\dfrac{\chi_0^{-1}(T)}{d\chi_0^{-1}(T)/dT}=\dfrac{(T-T _{\text{C}} )}{\gamma}
\end{equation}
\begin{figure*}
\includegraphics[width=\textwidth]{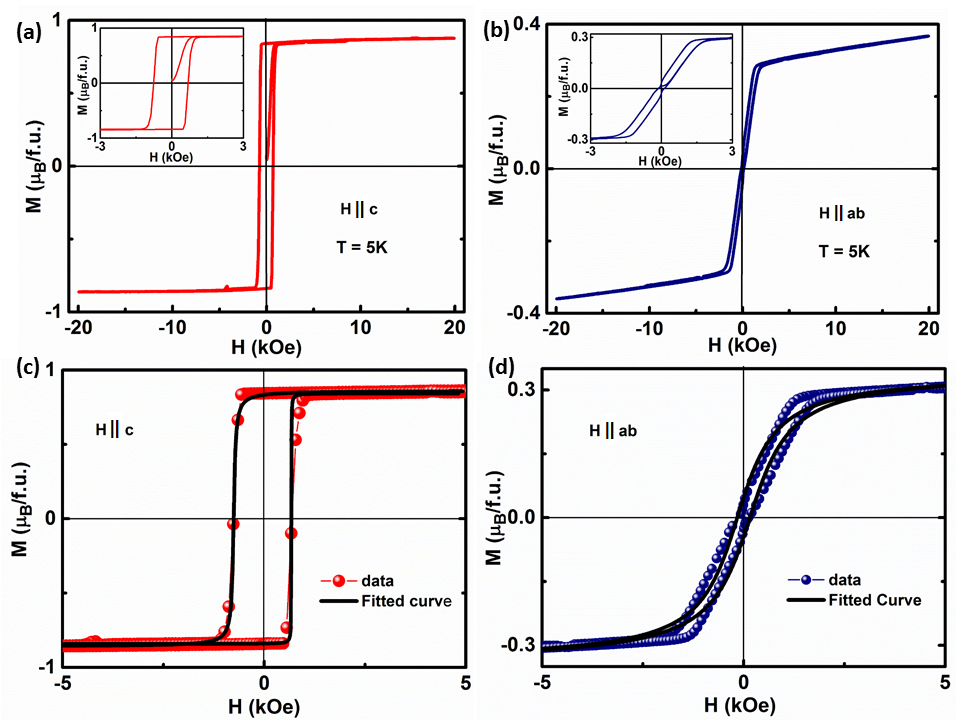}  
\caption{Isothermal magnetization $ M(H) $ under the sweeping magnetic field upto $ \pm20 $ kOe at 5 K in (a) $H\parallel c$  and (b) $H\parallel ab$ directions. The inset in (a) and (b) shows the zoomed view of magnetization curve at low magnetic fields showing a hysteresis loop in both the directions. Magnetic hysteresis loop is fitted according to Eq. (5) for (c) $H\parallel c$  and (d) $H\parallel ab$ directions in the range -5 kOe to +5 kOe.}
\end{figure*}
\par Figures 5(a) and 5(b) show the KF plots of both $ M_{\text{S}}(T)/(dM_{\text{S}}(T)/dT) $  vs. $ T $ and $ \chi_0^{-1}(T)/(d\chi_0^{-1}(T)/dT) $ vs. $ T $ for the directions $H\parallel c$ and $H\parallel ab$, respectively. In accordance to KF method, the KF plots described by Equations (3) and (4) are exactly linear whose respective slopes would yield $ 1/\beta $ and $ 1/\gamma $. In addition, $T _{\text{C}}  $ is determined from the intercept of the plots. Clear linear behaviour is observed in KF plots (figures 5(a) and 5(b)) in both the directions. The derived values of $T _{\text{C}}  $ and critical exponents are listed in Table I. It is found that the exponents $ \beta=0.307 $ and $ \gamma=1.15 $ in $H\parallel c$ direction are slightly less than the previous reports on Co$ _{3} $Sn$ _{2} $S$ _{2} $ \cite{65}. Here, the experimental values are approximately close to the predicted theoretical value for three dimensional (3D) Ising model $ (\beta=0.324, \gamma=1.241) $, which further indicates a 3D Ising type ferromagnetism in Co$ _{3} $Sn$ _{2} $S$ _{2} $. The difference in critical exponents along the two directions reflect the high anisotropy of the material. Furthermore, several factors including different magnetic domain structures, magnetic orientation or strain field effect could lead to the difference in the values obtained for two different crystallographic directions \cite{66}. 
\begin{table}
\caption{Extracted values of critical temperature $T _{\text{C}}  $, critical exponents $ (\beta,\gamma) $ from fitting to equations (3) and (4) in Kouvel-Fisher (KF) plots.}
\begin{ruledtabular}
\begin{tabular}{ c c c c c }  
 \textbf{Direction} & $ T_\text{C}^\# $ & $ \beta^{\#} $ & $ T_\text{C}^* $ & $\gamma^* $\\
& (K) & & (K)\\
\hline
$H\parallel c$ & 176.52 & 0.312 & 178.94 & 1.15 \\
$H\parallel ab$ & 177.18 & 0.022 & 175.29 & 0.567 \\
\end{tabular}
\end{ruledtabular}
\begin{flushleft}
$ \# $-KF plot of magnetization\\
$ * $-KF plot of inverse susceptibility
\end{flushleft}
\end{table}
\par Figures 6(a) and 6(b) show the magnetization $ M(H) $ as a function of magnetic field under the sweeping field from -20 kOe to +20 kOe taken at 5 K along $H\parallel c$ and $H\parallel ab$ directions, respectively. Initially, the $ M-H $ curves exhibit increase in magnetization at low magnetic fields and a pronounced hysteresis is also observed in both the directions. The inset in figures 6(a) and 6(b) depicts the enlarged image of $ M-H $ hysteresis loop at small magnetic fields. Similar hysteresis in two crystallographic directions have been reported earlier in Co$ _{3} $Sn$ _{2} $S$ _{2} $ \cite{10,14}. However, the magnetization saturates with the increase in magnetic field in $H\parallel c$ direction while it increases linearly with no signs of saturation in $H\parallel ab$ direction. The different hysteresis in two directions indicates the strong magnetic anisotropy i.e., large magnetocrystalline anisotropy present in our sample. Since the moments are saturated in c-axis direction, it is confirmed that the c-axis is the preferred easy axis of magnetization whereas the linear increase in isothermal magnetization along basal plane indicates that the hard axis of magnetization is along the ab-plane \cite{67}. The magnetic moments in $H\parallel ab$ quickly reach the value around 0.3$ \mu_{\text{B}} $/f.u. at low fields and thereafter, increase slowly in a linear fashion upto the highest applied magnetic field of 20 kOe. In other words, $ M-H $ curves in two directions exhibit a hysteresis switch from low field steep change to high field gradual change of the magnetization. This crossover  is  further accredited to magnetization reversal, which implies the presence of short range Co-Co magnetic interactions \cite{68,69}. In order to acquire more insight into the  hysteresis behaviour, the function in Eq. (5) is fitted to experimentally measured hysteresis loop in $H\parallel c$ and $H\parallel ab$ directions at 5 K (figures 6(c) and 6(d)) \cite{68,70}.
\begin{figure}
\begin{minipage}{0.5\textwidth}
\includegraphics[width=\textwidth]{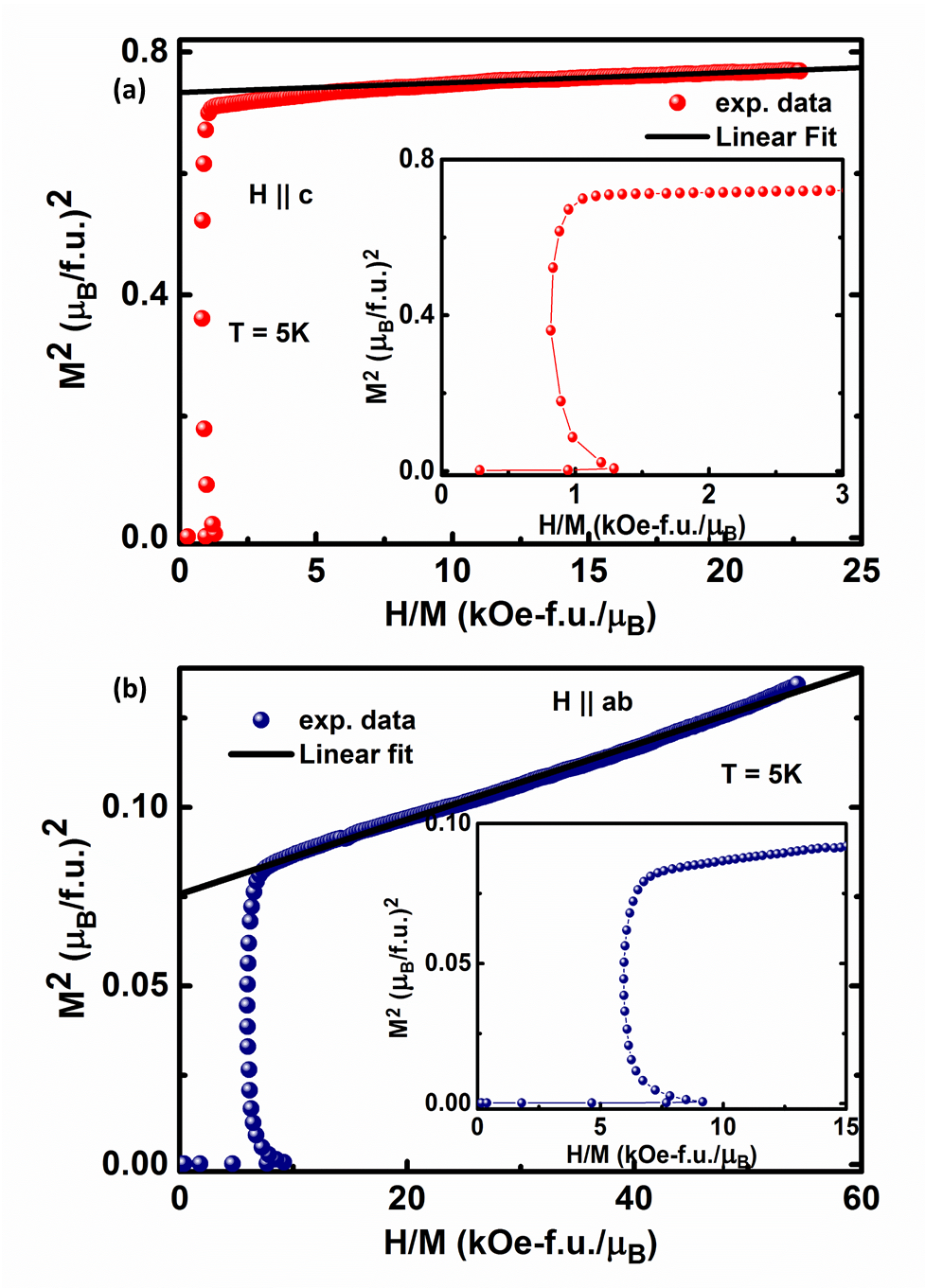} 
\caption{
Arrott plots ($ M^{2} $ vs. $ H/M $)  derived from isothermal magnetization curve in (a) $H\parallel c$ and $H\parallel ab$ directions. The black solid curve is the linear fitting to $ M^{2} $ data in the high field region. The inset in (a) and (b) shows the Arrott plot at low magnetic fields.}
\end{minipage}
\end{figure}
\begin{equation}
M(H)=\sum_{i=1}^{\infty}\dfrac{2M_{\text{S}}}{\pi}tan^{-1} \mid \dfrac{(H\pm H_{\text{C}})}{H_{C}}\tan (\dfrac{\pi \text{S}}{2}) \mid
\end{equation}
where $ M_{\text{S}} $ and $ H_{\text{C}} $ are the saturation magnetization and coercivity, respectively. Here, the fitted parameter $ S $ represents the ratio of remanent magnetization $ M_{\text{R}} $ and saturation magnetization i.e., $ S=M_{\text{R}}/M_{\text{S}} $. It is noted that the above function nicely fits to the measured data in both the directions. The fitted parameters obtained from the fitting function are listed in Table II. 
\begin{table}[b]
\caption{Parameter values of saturation magnetization $ M_{\text{S}} $, coercive field $ H_{\text{C}} $, ratio $ S=M_{\text{R}}/M_{\text{S}} $, remanent magnetization  $ M_{\text{R}} $ derived from fitting to Eq. (5) in magnetic hysteresis curves for two crystallographic directions.}
\begin{ruledtabular}
\begin{tabular}{ c c c c c }
 \textbf{Direction} & $ M_{\text{S}} $ & $ H_{\text{C}} $ & $ S=M_{\text{R}}/M_{\text{S}} $ & $ M_{\text{R}} $\\
& ($ \mu_{\text{B}} $/f.u.) & (kOe) & & ($ \mu_{\text{B}} $/f.u.)\\
\hline
$H\parallel c$ & 0.84 & 0.72 & 0.98 & 0.82 \\
$H\parallel ab$ & 0.35 & 0.15 & 0.12 & 0.042 \\
\end{tabular}
\end{ruledtabular}
\end{table}
\par It is mentioned that the low coercivity has been observed in $H\parallel ab$  direction as compared to $H\parallel c$ direction. Moreover, the smaller $ M_{\text{R}}/M_{\text{S}} $  ratio in $H\parallel ab$ direction in contrast to $H\parallel c$ direction suggests the enhanced disoriented ferromagnetic domains \cite{71}. Clearly as evident from the figures, the hysteresis loop observed along $H\parallel c$ direction (figure 6(a) inset) reflects the hard magnetic phase revealing high coercivity and magnetic saturation. On the other hand, the hysteresis loop in $H\parallel ab$ direction shown in the inset of figure 6(b) depicts the soft magnetic phase that exhibits a low coercive field without any saturation at higher applied magnetic fields. It is inferred that the magnetic saturation in hard magnetic phase arises due to contribution of FM phase whereas the soft magnetic phase (where no saturation is observed) arises due to the competition between ferromagnetic and antiferromagnetic interactions.
Next, we present the Arrott plots ($ M^{2} $ vs. $ H/M $) at 5 K for $H\parallel c$ and $H\parallel ab$ directions  in figures 7(a) and 7(b), respectively. It is noticed that the curves exhibit a high field or linear behaviour with a strong downward curvature at very low magnetic fields (insets of figures 7(a) and 7(b)). Similar convex behaviour in $ M^{2} $ is reported in typical itinerant ferromagnetic materials such as MnSi \cite{23} and transition metal oxide Sr$ _{1.5} $Nd$ _{0.5} $MnO$ _{4} $ \citep{72}. According to Banerjee’s criterion, the convex downward curvature observed along the two directions resembles a second order phase transition \cite{73}. It is also noteworthy here that the linear positive slope in Arrott plot indicates a second order FM to PM phase transition in Co$ _{3} $Sn$ _{2} $S$ _{2} $.
\par To get further deeper insights into the presence of FM clusters, the isothermal magnetization curves over the temperature range above and below $ T_{\text{C}} $ under the sweeping magnetic field of 20 kOe in $H\parallel c$ are depicted in figure 8(a). The $ M-H $ curves at different temperatures exhibit a non-linear behaviour as depicted in the inset of figure 8(a) implying that the magnetic state above $ T_{\text{C}} $ and below $ T_{\text{GP}} $ is not a pure PM state and validates the GP owing to the presence of short range FM clusters \cite{74}. Moreover, the isothermal magnetization data for temperatures above $ T_{\text{C}} $ are also analysed using the Arrott plot as shown in figure 8(b). The inset in figure 8(b) presents the zoomed part of Arrott plot analysed in the temperature range ($ T_{\text{C}}<T<T_{\text{GP}}$) upto $ \pm 10 $ kOe. It is clearly observed that the linear extrapolation of Arrott plot curves on $ M^{2} $ axis yields no spontaneous magnetization $ M_{\text{sp}} $ at all temperatures above $ T_{\text{C}} $. The zero $ M_{\text{sp}} $ obtained between $ T_{\text{C}} $ and $ T_{\text{GP}} $ again confirms that the magnetic state above $ T_{\text{C}} $ is GP and not a pure PM state, and hence validates the presence of FM clusters in Co$ _{3} $Sn$ _{2} $S$ _{2} $ \cite{22}.
\begin{figure*}
\includegraphics[width=\textwidth]{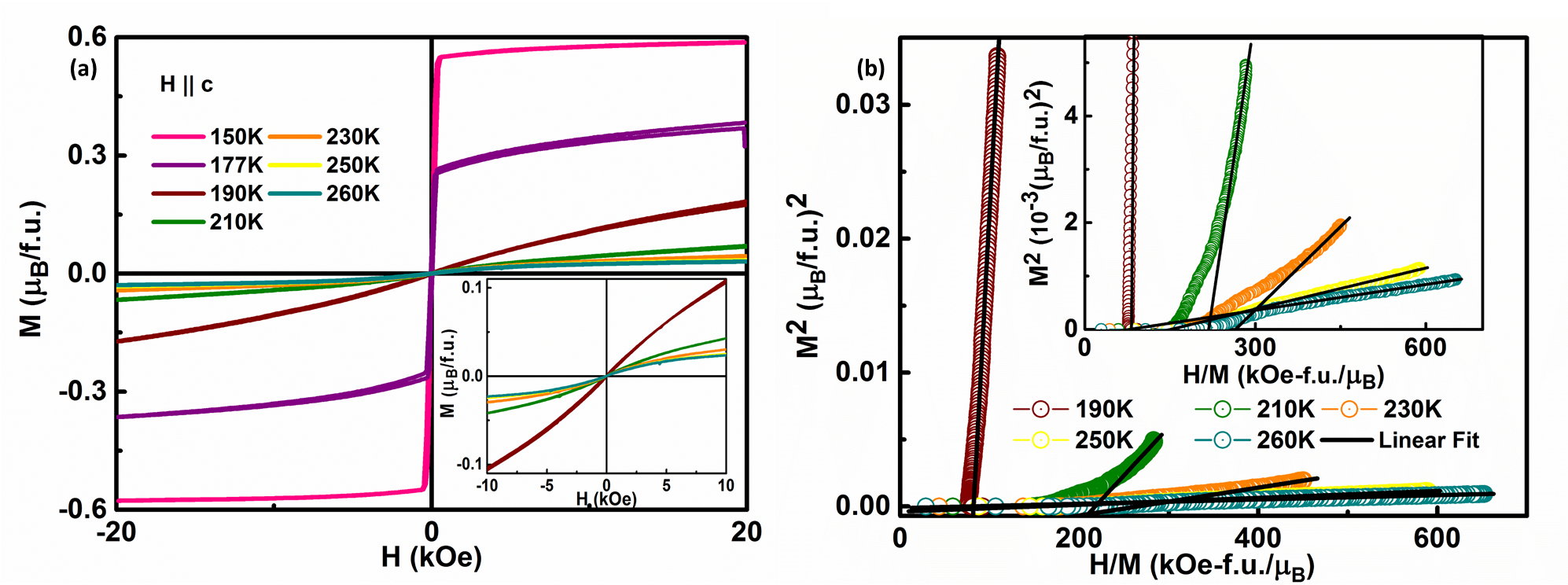} 
\caption{(a) The plot of M-H curves at different temperatures above and below $ T_{\text{C}} $ in the applied magnetic field from -20 kOe to +20 kOe in $H\parallel c$ direction. Inset shows the M-H curves upto $ \pm 10 $ kOe in the temperature range ($ T_{\text{C}}<T<T_{\text{GP}}$). (b) The plot of $ M^{2} $ vs. $ H/M $ for the temperatures ranging between $ T_{\text{C}} $ and $ T_{\text{GP}} $. }
\end{figure*}
\par The characteristic feature of non-linearity and convex curvature in the Arrott plot (figures )at extremely low fields is explained by means of Takahashi spin fluctuation theory \cite{23}. In Takahashi spin fluctuation theory, 
the magnetization at the phase transition is given by the following expression \cite{23,75,76}:
\begin{equation}
H=\dfrac{F_{1}}{N_0^3(g\mu_{\text{B}})^{4}}\times (-M_0^2 + M^{2})
\end{equation}
where $ H $ is the magnetic field, $ N_{0} $ is the Avogadro number, $ g $ is the Lande g-factor, $ F_{1} $ is the mode-mode coupling term and $ M_{0} $ is the spontaneous magnetization. 
$ F_{1} $ can be estimated experimentally at low temperatures from the slope of Arrott plots ($ M^{2} $ vs. $ H/M $) using Eq. (7)\cite{77}:
\begin{equation}
F_{1}=\dfrac{N_0^3(2\mu_{\text{B}})^{4}}{k_{\text{B}}\zeta}
\end{equation}
where $ \mu_{\text{B}} $ is the Boltzmann constant and $ \zeta $ is the slope of Arrott plot curve at a desired temperature. The spin fluctuation parameters $ T_{0} $ and $ T_{\text{A}} $ indicate the distribution widths in the energy and wave-vector spaces, respectively. These parameters are related to $ F_{1} $ by the following equations:
\begin{equation}
(\dfrac{T _{\text{C}} }{T_{0}})^{5/6}=\dfrac{p_\text{s}^2}{5g^{2}C_{4/3}}(\dfrac{15cF_{1}}{2T_{\text{C}}})^{1/2}
\end{equation}
\begin{equation}
(\dfrac{T _{\text{C}} }{T_{\text{A}}})^{5/3}=\dfrac{p_s^2}{5g^{2}C_{4/3}}(\dfrac{2T_{C}}{15cF_{1}})^{1/2}
\end{equation}
where $ C_{4/3}=1.00608 $ and $ p_{\text{s}} $ are the spontaneous magnetic moment. Takahashi theory ensures that the spin fluctuation parameters can be estimated from measured magnetic data only. The fraction $ T _{\text{C}} /T_{0} $  in the spin fluctuation theory specifies the degree of itinerant character of magnetic atoms with a strong itinerant feature achieved at $ T _{\text{C}} \ll T_{0} $. Besides, $ T _{\text{C}} /T_{0} $  is related to the Rhodes-Wohlfarth ratio $ p_{\text{c}}/p_{\text{s}} $ in Takahashi theory, stating the large $ p_{c}/p_{s} $ and small $ T _{\text{C}} /T_{0} $  indicative of weak ferromagnetism. Here, $ p_{\text{c}}=\sqrt{(1+p_{\text{eff}}^2)}-1 $ is the magnetic moment, in units of $ \mu_{\text{B}} $, in the paramagnetic phase and $ p_{\text{eff}} $ is the effective magnetic moment calculated from Curie-Weiss law. Such itinerant ferromagnetism principally arises from the screening effect of electrons in the band structure due to electron-electron correlations and exchange interactions \cite{72}.It is pertinent to note that the factors $ p_{\text{c}}/p_{\text{s}} $  and $ T _{\text{C}} /T_{0} $  are the important parameters for characterising the degree of itinerancy of electrons \cite{78}.
Now using the linearly fitted slope $ \zeta $ and intercept values in the Arrott plot of $ M^{2} $ vs. $ H/M $ in figures 7(a) and 7(b), we can estimate the parameters coupling term $ F_{1} $ and the spontaneous magnetic moment $ p_{\text{s}} $, respectively. 
The derived values of $ F_{1} $, $ p_{\text{s}} $ and $ p_{\text{eff}} $ for both the directions are listed in the Table III. 
 The spin fluctuation parameters $ T_{\text{A}} $ and $ T_{0} $ are evaluated from $ p_{\text{s}} $, $ F_{1} $ and $T _{\text{C}}  $ values using the relations given in equations (8) and (9), respectively 
are also listed in the Table III. 
The ratios $ p_{\text{c}}/p_{\text{s}} $  and $ T _{\text{C}} /T_{0} $  have been estimated and also listed in the Table III. As observed, the factor $ p_{\text{c}}/p_{\text{s}} $  is found reasonably large (i.e, $ p_{\text{c}}/p_{\text{s}} > 1$) in both the directions. Moreover, $ T _{\text{C}} /T_{0} $  is determined to be very small i.e., less than unity. Therefore, a large $ p_{\text{c}}/p_{\text{s}} $  and small $  T _{\text{C}} /T_{0} $ values found along the two directions clearly imply the itinerant nature of ferromagnetism in Co$ _{3} $Sn$ _{2} $S$ _{2} $. 
The discrepancy in the two ratios in the directons again indicates the anisotropy present in Co$ _{3} $Sn$ _{2} $S$ _{2} $. It is found that the respective values of $ T _{\text{C}} /T_{0} $ in $H\parallel c$ and $H\parallel ab$ directions are approximately $0.24$ and $0.022$, which means a strong itinerant character of Co$ _{3} $Sn$ _{2} $S$ _{2} $ \cite{79}. 
\begin{table*}
\caption{Spontaneous magnetic moment $ p_{\text{s}} $, basic spin fluctuation parameters $ T _{\text{C}} , F_{1}, T_{\text{A}}, T_{0}$ obtained from fitting in Arrott plot in the two field directions on the basis of Takahashi’s spin fluctuation theory. Effective magnetic moment $ p_{\text{eff}} $ is calculated from Curie-Weiss fitting in the inverse susceptibility data.}
\begin{ruledtabular}
\begin{tabular}{ c c c c c c c c c c }
 \textbf{Direction} & $ p_{\text{s}} $ & $ p_{\text{eff}} $ & $T _{\text{C}}  $ & $ F_{1} $ & $ T_{\text{A}} $ & $ T_{0} $ & $ T _{\text{C}} /T_{0} $ & $ p_{\text{C}} $ & $ p_{\text{C}}/p_{\text{s}} $\\
& ($ \mu_{\text{B}} $/f.u.) & ($ \mu_{\text{B}} $/f.u.) & (K) & (K) & (K) & (K) & & ($ \mu_{\text{B}} $/f.u.) \\
\hline
$H\parallel c$ & 0.86 & 2.77 & 177 & 722 & 660 & 1.74$ \times 10^{3} $ & 0.25 & 1.95 & 2.26 \\
$H\parallel ab$ & 0.28 & 6.66 & 177 &  1.031$ \times 10^{3} $ & 1.732$ \times 10^{3} $ & 8.13$ \times 10^{3} $ & 0.022 & 5.74 & 20.48 \\
\end{tabular}
\end{ruledtabular}                                                                                                                                                                                                                                                                                                                                                                                                                                                                                                                                                                                                                                                                              
\end{table*}
\par Figure 9 shows the $ M-H $ curve fitted with Brillouin function given in Eq. (10) at 5 K for $H\parallel c$ and $H\parallel ab$ directions. In general, the magnetization of a FM system can be defined by the Brillouin function \cite{80}.
\begin{equation}
M(x)=NgS\mu_{\text{B}}[\dfrac{2S+1}{2S}\coth (\dfrac{2S+1}{2S})x-\dfrac{1}{2S}\coth (\dfrac{x}{2S})]
\end{equation}
where $ x=\frac{gS\mu_{\text{B}}H}{k_{\text{B}}T} $ is the ratio of Zeeman energy of the magnetic moment in molecular field thermal energy, $ g $ is the Lande $ g $-factor value and $ N $ is the number of spins. The lower and higher spin state values of Co$ ^{2+} $ ions i.e., $ S=1/2 $ and $ S=3/2 $ are taken into account for the analysis using Brillouin function. It is noted for both the values of $ S $, the function fit well at low and high magnetic fields in the two directions.
\begin{figure}[h]
\begin{minipage}{0.45\textwidth}
\includegraphics[width=\textwidth]{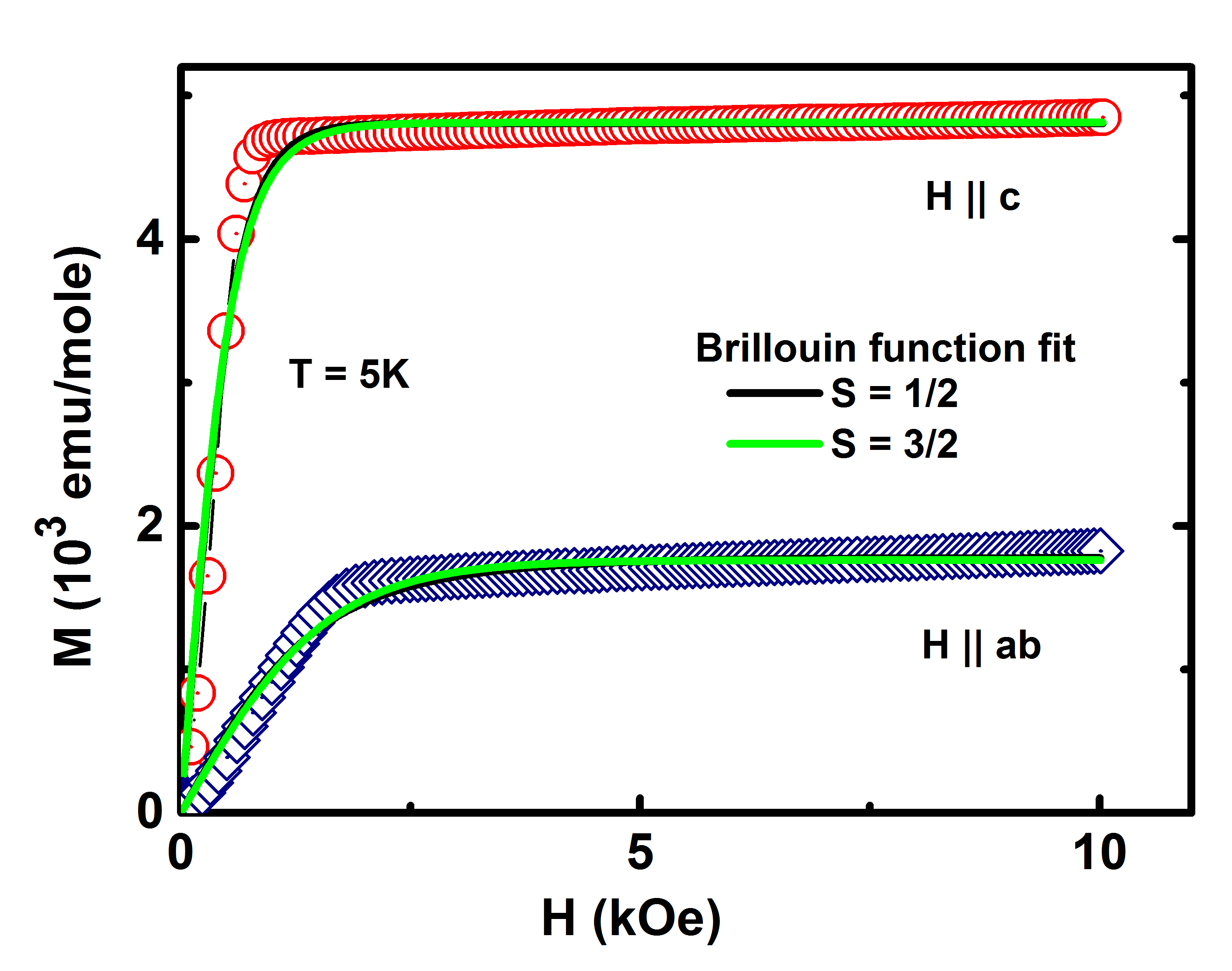} 
\caption{Magnetic field dependent magnetization $ M(H) $ at 5 K is fitted with Eq. (10) for $H\parallel c$ and $H\parallel ab$ directions. The black line correspond to the fitting curves for the spin values $ S=1/2 $ (black curve) and $ S=3/2 $ (green curve).}
\end{minipage}
\end{figure}
\par The magnetic field dependent magnetization data taken at 1.7 K is further analysed to extract the information about magneto-crystalline anisotropy in Co$ _{3} $Sn$ _{2} $S$ _{2} $. The magnetocrystalline anisotropy is related to the ground state energy required to orient the magnetization vector from easy magnetic axis to other random directions. The effective magnetocrystalline anisotropy constant $ K_{\text{eff}} $ is determined from the relation: $ K_{\text{eff}}=M_{\text{S}}H_{\text{a}}/2 $. The parameter $ H_{\text{a}} $ is the anisotropy field defined as that particular magnetic field where the extrapolated lines of magnetization curves from the out-of plane and in-plane directions intersect \cite{67}. It is clearly shown by a pointed arrow in figure 10 that the linear extrapolation of $ M-H $ curve along \textit{ab}-plane reaches the saturation magnetization in \textit{c}-axis direction at a magnetic field $ H_{\text{a}} \sim $ 185 kOe. Hence, the magnitude of anisotropy constant $ K_{\text{eff}} $ is evaluated to be about 9.24 $ \times 10^{6} $ erg/cm$ ^{3} $, which is quite comparable to the previous result in Co$ _{3} $Sn$ _{2} $S$ _{2} $ \cite{67}. Further, the obtained $ K_{\text{eff}} $ value is found to be quite larger than the values reported in other known kagome ferromagnetic materials such as CrI$ _{3} $ \cite{81}, Fe$ _{3} $GeTe$ _{2} $ \cite{82} and Fe$ _{3} $Sn$ _{2} $ \cite{83}.
\begin{figure}
\begin{minipage}{0.45\textwidth}
\includegraphics[width=\textwidth]{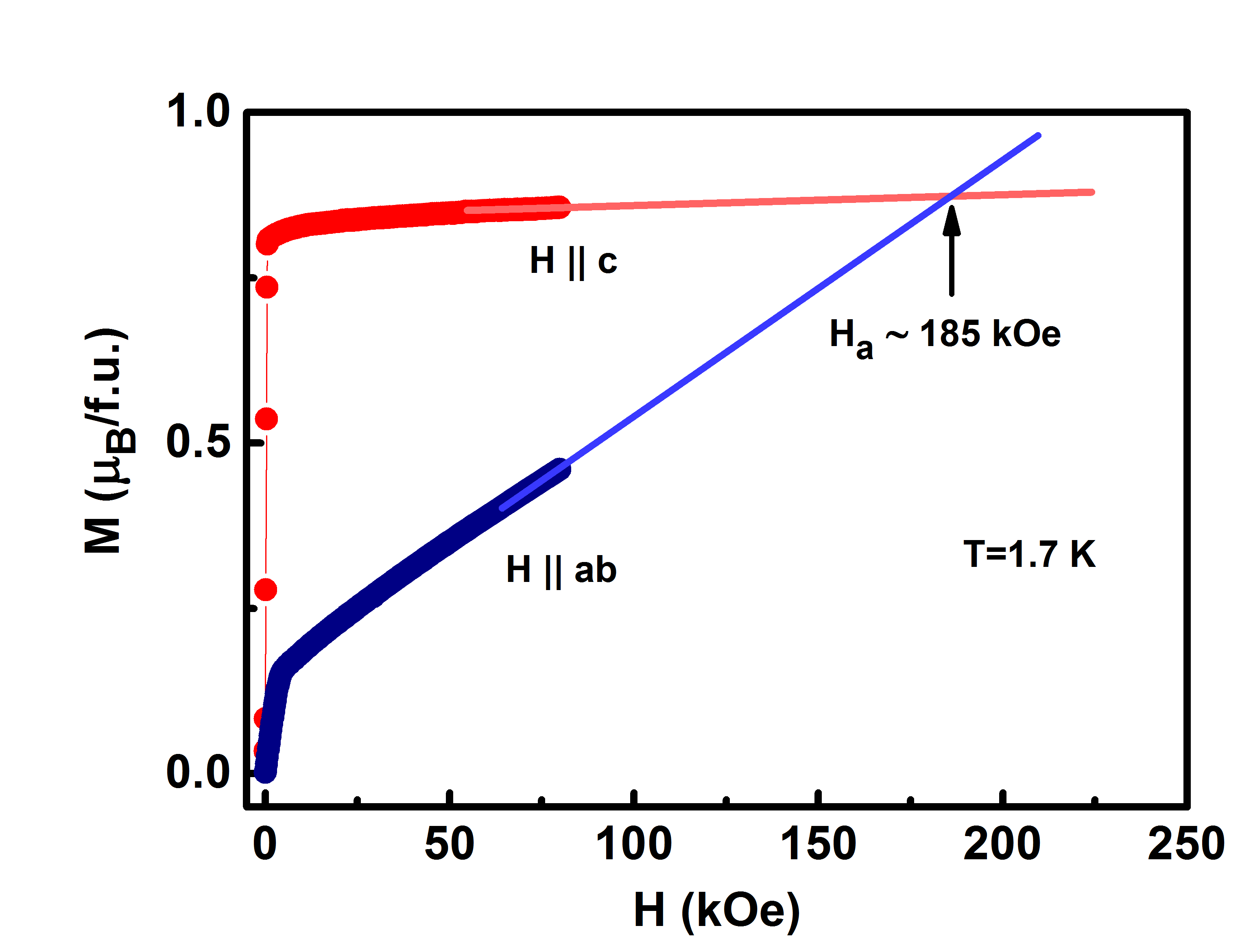} 
\caption{The plot of linear extrapolation of $ M(H) $ curves at 5 K along $H\parallel c$ and $H\parallel ab$ directions intersect at a magnetic field to determine the anisotropy field indicated by a pointed arrow. }
\end{minipage}
\end{figure}
It is understood that the large magnetocrystalline anisotropy is accompanied by high anisotropy field. Thus, in Co-based compounds, strong magneto-crystalline anisotropy in magnetization is predominantly due to the large orbital magnetic moments concomitant with Co atoms coupled with the crystal electric field effects in the crystal structure \cite{84,85}. Such large magnetocrystalline anisotropy originates from strong spin-orbit coupling effect at the lattice sites in Co$ _{3} $Sn$ _{2} $S$ _{2} $ \cite{28}.

\section{\label{sec:level4}Conclusion}
In conclusion, we have demonstrated the evidence of ferromagnetic short-range clusters in the paramagnetic region along with low temperature cluster-glassy phase in highly anisotropic kagome compound Co$ _{3} $Sn$ _{2} $S$ _{2} $. The presence of short range ferromagnetic clusters is demonstrated by a sharp downturn in the paramagnetic region in the inverse magnetic susceptibility data above $T _{\text{C}} $, showing a clear deviation from linear Curie-Weiss behaviour. This feature is a characteristic signature of pronounced Griffiths phase in Co$ _{3} $Sn$ _{2} $S$ _{2} $. In addition, a strong Griffiths singularity is observed at the temperature around 265 K followed by the slow magnetic relaxation behaviour and no net spontaneous magnetization above $T _{\text{C}} $ in the Griffiths phase. The crossover, from low field steep increase to high field gradual change of magnetization, observed in magnetic hysteresis further corroborates the short range magnetic interactions present in the system. The itinerant ferromagnetism in Co$ _{3} $Sn$ _{2} $S$ _{2} $ is confirmed from Arrott plots as established on the basis of Takahashi theory of spin fluctuations. A large magnetocrystalline anisotropy is found which is related to the high anisotropy field, suggesting the role of strong spin-orbit coupling in Co$ _{3} $Sn$ _{2} $S$ _{2} $. 
 \section*{Acknowledgement}
This work was acknowledged by DST-FIST, DST-PURSE and DST-SERB project under Grant No. PHY/2016/003998. Authors are thankful to AIRF, JNU for the PPMS facilities. 

\end{document}